\documentstyle[12pt,epsf,rotate]{article}
\def\MeV{\,{\rm MeV}}

\def\Mpc{\,{\rm Mpc}}

\def\la{\mathrel{\mathpalette\fun <}}
\def\ga{\mathrel{\mathpalette\fun >}}
\def\fun#1#2{\lower3.6pt\vbox{\baselineskip0pt\lineskip.9pt
  \ialign{$\mathsurround=0pt#1\hfil##\hfil$\crcr#2\crcr\sim\crcr}}}

\begin{document}
\pagestyle{empty}
\begin{center}
\bigskip

\bigskip
%\rightline{7 June 1996}
\medskip
\rightline{FERMILAB--Pub--96/122-A}
\rightline{astro-ph/9606059}
\rightline{submitted to {\it Physical Review Letters}}

\vspace{.2in}
{\Large \bf The Big-Bang Nucleosynthesis Limit to the Number
of Neutrino Species}
\bigskip

\vspace{.2in}
Craig J.~Copi,$^{1,2}$ David N.~Schramm,$^{1,2,3}$ and
Michael S. Turner$^{1,2,3}$\\

\vspace{.2in}
{\it $^1$Department of Physics \\
Enrico Fermi Institute, The University of Chicago, Chicago, IL~~60637-1433}\\

\vspace{0.1in}
{\it $^2$NASA/Fermilab Astrophysics Center\\
Fermi National Accelerator Laboratory, Batavia, IL~~60510-0500}\\

\vspace{0.1in}
{\it $^3$Department of Astronomy \& Astrophysics\\
The University of Chicago, Chicago, IL~~60637-1433}\\

\end{center}

\vspace{.3in}
\centerline{\bf ABSTRACT}
\bigskip

Concern about systematic uncertainty in the $^4$He abundance as well
as the chemical evolution of $^3$He leads us to re-examine
this important limit.  It is shown that with conservative assumptions
no more than the equivalent of 4 massless neutrino species are allowed.
Even with the most extreme estimates of the astrophysical uncertainties
a meaningful limit still exists, less than 5 massless neutrino
species, and illustrates the robustness of this argument.
A definitive measurement of the deuterium abundance in high-redshift
hydrogen clouds should soon sharpen the limit.

\newpage
\pagestyle{plain}
\setcounter{page}{1}
\newpage

{\it Introduction.}  Big-bang nucleosynthesis is
one of the experimental pillars of the standard cosmology
\cite{cst1,bbnrev}.  It also probes particle physics.
Among other things, it has been used to set a stringent
limit to the energy density contributed by light (mass $\ll \MeV$)
particle species, usually quantified as the equivalent number of
massless neutrino species ($\equiv N_\nu$)
\cite{nulimit}.  This limit indicated that the number of neutrino
species was small before accelerator experiments were able to
experimentally establish this directly, and further, has served to
constraint theories proposed to unify the forces and
particles of Nature.  Since the cosmological bound
also constrains light species that do not couple to
the $Z^0$, it is an important complement to the LEP measurement,
$N_\nu=2.991\pm 0.016$.

The physics underlying the neutrino limit is simple:  the big-bang
production of $^4$He increases with both baryon density
(quantified by the present baryon to photon ratio $\eta$)
and the number of massless neutrino species.  Thus, an upper limit
to the primeval $^4$He abundance ($\equiv Y_P$) and a lower limit to the
baryon density lead to an upper limit to $N_\nu$.
The lower limit to the baryon density is based upon the big-bang production of
deuterium, which rises rapidly with decreasing baryon density \cite{yangetal}.
Because D is easily destroyed in stars, first being burnt
to $^3$He, the limit actually hinges upon
D + $^3$He, bringing in the chemical evolution of $^3$He.
This has been the standard approach for setting a limit to $N_\nu$.

Over the past five years limits to $N_\nu$ ranging from 3.04 to around 5 have
been quoted \cite{recentnulimits}.  The disparity arises
because the irreducible uncertainties are systematic,
rather than statistical.  In particular, the pressing issues
involve the primeval abundance
of $^4$He and the chemical evolution of $^3$He (astronomers
use the term ``chemical evolution'' to refer to nuclear processing).
The purpose of this paper is to clarify the current situation,
to use a new technique to obtain a limit to $N_\nu$ that is
independent of the chemical evolution of $^3$He, and to
show how measurements of the primeval D abundance in
high-redshift hydrogen clouds should soon sharpen the neutrino limit.
In so doing, we will emphasize the robustness of the cosmological
limit to $N_\nu$.

{\it The light-element abundances.}  To orient the reader we
begin with a brief overview.  The big-bang production of
D, $^3$He, $^4$He and $^7$Li is
summarized in Fig.~1.  The predicted and measured abundances are
consistent---within the uncertainties---for $\eta
\simeq (2 - 7)\times 10^{-10}$ \cite{cst1,right}.  The status
of the four light elements is as follows.

\begin{enumerate}

\item The lithium abundance is measured in the atmospheres of the oldest
stars in our galaxy (pop II halo stars) and is relatively
well determined, ($^7$Li/H)$_{\rm pop\ II} = (1.4\pm 0.3)\times
10^{-10}$ \cite{cst1,thorburn}.  Because $^7$Li burns at a low temperature,
the most important concern is that $^7$Li may have been depleted.
Stellar models indicate that a depletion of up to a
factor of about two is consistent with the
observations of $^7$Li and other light elements in
these old stars \cite{li7deplete}.

\item The $^4$He abundance increases with time, as $^4$He is made by stars,
and so any measurement provides an upper
bound to the primeval component.  The primeval abundance
is extrapolated from measurements of
metal-poor, extragalactic HII regions (hot clouds of
ionized H and He gas).  A compilation and analysis of the
extant data gave $Y_P = 0.232 \pm 0.003{\rm\,(stat)} \pm 0.005{\rm\,(sys)}$
\cite{olivesteigman}.  Others have estimated the systematic
uncertainty to be larger, by a factor of two or even more \cite{he4sys};
it arises from many sources, from uncertainties in
theoretical emissivities to assumptions and approximations made
in modeling the HII regions.  A very recent study of 27 new extragalactic
HII regions, using a new set of emissivities, finds $Y_P =
0.243\pm 0.003$ \cite{izotov96}.

\item Deuterium is destroyed in the contemporary Universe,
being burnt to $^3$He, which links its post-big-bang history
to that of $^3$He.  The deuterium abundance has been measured
in the local interstellar medium (ISM) and in the pre-solar nebula \cite{dlocal}.
While neither value is expected to reflect the primeval abundance, both
place a lower limit to the big-bang production and thereby
a firm upper limit to the baryon density.  This two-decade old
limit, $\Omega_B \la 0.06\,(70\,{\rm km}\,{\rm sec}^{-1}\Mpc^{-1}/H_0)^2$,
precludes baryons from closing the Universe \cite{reevesetal},
and is not being questioned.

The primeval deuterium abundance can be determined
by measuring the deuterium Ly-$\alpha$ absorption feature in high-redshift ($z\ga
2$) hydrogen clouds which are backlit by distant
QSOs \cite{adams,newsviews}.  There are now three claimed detections
\cite{detect} and four tentative detections \cite{maybedetect}; the reported
values are from (D/H)$_P \simeq 2\times 10^{-5}$ to $2\times 10^{-4}$,
spanning the range anticipated on the basis of the big-bang production
of the other light-elements.  A definitive determination will peg the
baryon-to-photon ratio to an accuracy of around 10\% because of the
rapid variation of D production with $\eta$.

\item Since all measurements of $^3$He involve samples that
have to some extent been processed through stars,
chemical evolution is the central issue.  The chemical evolution
of $^3$He is complicated and not well understood.
Stars burn D to $^3$He before reaching the main
sequence; this is not controversial.  According to conventional stellar
models, massive stars ($M\ga 8M_\odot$) reduce the $^3$He
in material they return to the ISM, by a factor of around five or so.
Since high-mass stars also produce elements heavier than $^4$He
(``metals''), there is a limit to the amount of processing through
high-mass stars -- and thus to $^3$He destruction.  
Low-mass stars ($M\la 2M_\odot$) preserve and add significant
$^3$He to the material they return to the ISM \cite{he3conv}.
This leads to
the conclusion that the sum of D + $^3$He should have decreased by
at most a modest factor since primordial nucleosynthesis,
which has been used to obtain the lower bound to $\eta$ needed to constrain
the number of massless neutrino species \cite{yangetal}.

There is little empirical evidence for the conventional view concerning
the chemical evolution of $^3$He; moreover, it
has recently  been challenged both theoretically and observationally.
Slow-mixing mechanisms that would transport $^3$He deep enough to be burnt
have been suggested to explain oxygen and carbon isotopic ratios
measured in some stars and meteorites \cite{slowmix}.
In addition, even if the conventional wisdom is correct in the
mean, galactic chemical abundances are heterogeneous and
the D + $^3$He abundance has only been measured locally \cite{cst2}.

On the observational side, a recent measurement of the $^3$He abundance in
the ISM indicates that the sum of D + $^3$He has not increased in
the past 4.5 Gyr \cite{gg,ttsc}; conventional models predict an increase
since the recent evolution of the ISM should be dominated by
low-mass stars.  Finally,
if the highest values of the D abundance measured in high-redshift
hydrogen clouds are correct, then the present D + $^3$He abundance
is too small by a factor of almost ten to accommodate conventional models.

\end{enumerate}

{\it A new approach.}  The two major obstacles
to setting a reliable limit to $N_\nu$ are
the chemical evolution of $^3$He -- which affects the lower bound to $\eta$
based upon D + $^3$He -- and the systematic uncertainty in the
primeval $^4$He -- which affects the upper bound to big-bang
$^4$He production.  To clarify and to minimize
the dependence upon these we have computed limits to $N_\nu$ based upon
a likelihood function for ${\tilde N}\equiv N_\nu -
\Delta Y/0.016$ which uses the
$^7$Li and $^4$He abundances\footnote{The authors of Ref.~\cite{fieldsolive}
have advocated the use of $^7$Li and $^4$He alone, albeit
in a different context.}  and takes deuterium into account only
by means of a Bayesian prior.  This approach exploits the fact
that $^7$Li increases with $N_\nu$ rapidly enough to provide
a meaningful lower limit to $\eta$ as well as the fact that $N_\nu$
and a systematic shift ($\Delta Y$) in the assumed primeval $^4$He
abundance are equivalent (see Ref.~\cite{right}).

In computing ${\cal L}(\tilde N)$ we take $Y_P = 0.242
+\Delta Y \pm 0.003$ and consider two possibilities
for $^7$Li, ($^7$Li/H)$_P =
(1.4\pm 0.3) \times 10^{-10}$ (no depletion) and $(3\pm 0.6)\times
10^{-10}$ (factor of two depletion), using whichever gives the
less stringent limit to $N_\nu$.
Regarding the primeval $^4$He abundance; as a
``standard case'' we take the central value $Y_P = 0.242$,
0.01 larger than the central value recommended in Ref.~\cite{olivesteigman} --
essentially equal to the value derived from the newest data set
\cite{izotov96} -- and roughly equal to the upper limit to $Y_P$ used
previously in deriving a limit to $N_\nu$.  Through $\Delta Y$
we allow for possible systematic error in the $^4$He abundance
and explore the sensitivity of the neutrino limit to it.

We consider four priors for deuterium:
(1) (D/H)$_P\le 1.0$, corresponding to
essentially no prior information; (2) (D/H)$_P\le 2\times 10^{-4}$,
corresponding to the previous bound based upon D + $^3$He \cite{yangetal};
(3) a distribution for (D/H)$_P$ based upon an extreme
chemical evolution model \cite{ttsc}, which allows for destruction of
$^3$He in low-mass stars and extra destruction in high-mass stars
by assuming that metals they produce
were ejected into the intergalactic medium,
avoiding the Galactic metallicity constraint;
and (4) (D/H)$_P = (2.5\pm 0.75)\times 10^{-5}$, an example
of the accuracy to which the primeval deuterium might
be determined.  Our four likelihood functions are shown in Fig.~2.

\begin{table} \center
\begin{tabular}{cccc} \hline
D/H ($^7$Li)   & $Y_P$ & $N_\nu\ge2$ & $N_\nu\ge3$ \\ \hline \hline
& $0.232\pm 0.003$ & 2.0-2.6, 2.8-3.6 & 3.0-3.7 \\
& $0.237\pm 0.003$ & 2.0-2.9, 3.2-3.9 & 3.0-4.0 \\
$\le 1.0 $ (high) &$0.242\pm 0.003$ &\bf 2.3-3.3, 3.4-4.3 &\bf 3.0-3.3, 3.4-4.3 \\
& $0.247\pm 0.003$ & 2.6-3.6, 3.8-4.6 & 3.0-3.6, 3.7-4.6 \\
& $0.252\pm 0.003$ & 2.9-3.9, 4.1-4.9 & 3.0-3.9, 4.1-4.9 \\ \hline
& $0.232\pm 0.003$ & 2.0-3.1 & 3.0-3.3 \\
& $0.237\pm 0.003$ & 2.3-3.4 & 3.0-3.5 \\
$<2\times10^{-4}$ (low)  & $0.242\pm 0.003$ &\bf 2.6-3.7 &\bf 3.0-3.7 \\
& $0.247\pm 0.003$ & 2.9-4.1 & 3.0-4.0 \\
& $0.252\pm 0.003$ & 3.2-4.4 & 3.2-4.4 \\ \hline
& $0.232\pm 0.003$ & 2.0-2.6 & 3.0-3.5 \\
& $0.237\pm 0.003$ & 2.1-2.9 & 3.0-3.6 \\
extreme chemical & $0.242\pm 0.003$ &\bf 2.4-3.2 &\bf 3.0-3.4 \\
evolution for $^3$He (high) & $0.247\pm 0.003$ & 2.6-3.7 & 3.0-3.6 \\
& $0.252\pm 0.003$ & 2.9-4.0 & 3.0-3.8 \\ \hline
& $0.232\pm 0.003$ & 2.0-2.4 & 3.0-3.1 \\
& $0.237\pm 0.003$ & 2.0-2.7 & 3.0-3.2 \\
$(2.5\pm 0.75)\times10^{-5}$ (high)& $0.242\pm 0.003$ &\bf 2.3-3.0 &\bf 3.0-3.2 \\
& $0.247\pm 0.003$ & 2.6-3.3 & 3.0-3.4 \\
& $0.252\pm 0.003$ & 2.9-3.6 & 3.0-3.6 \\
\hline
\hline
\end{tabular}
\caption{The 95\% credible intervals for $N_\nu$ based upon
the different deuterium and $N_\nu$
priors and central values of the primeval $^4$He abundance.
High (low) indicates high (low) $^7$Li abundance was used to derive
the limit (see text).}
\end{table}

{\it Discussion.}  Credibility intervals for
$N_\nu$ are obtained from ${\cal L}(\tilde N)$ by choosing a value for $\Delta
Y$ and imposing a prior on $N_\nu$, either $N_\nu \ge 3$
or $N_\nu \ge 2$.  As Table 1 illustrates, better knowledge of the
deuterium abundance and/or a tighter limit to the primeval
$^4$He abundance improve the limit to $N_\nu$.

Because our likelihood is a
function of ${\tilde N} = N_\nu -\Delta Y/0.016$, in the
absence of a prior for $N_\nu$ the limit to $N_\nu$
would scale precisely as the limit to $\tilde N$
plus $\Delta Y/0.016$; e.g., increasing $Y_P$ by 0.01 would raise
the limit by about 0.6 neutrino species.  This scaling works
reasonably well for the $N_\nu \ge 2$ prior.

Regarding the dependence of the neutrino limit upon deuterium:

\begin{enumerate}

\item Using the original D + $^3$He bound, (D/H)$_P<2\times 10^{-4}$
\cite {yangetal}, the 95\%
credible region extends to ${\tilde N} = 3.3$.  For
$Y_P = 0.242$, $N_\nu < 3.7 (3.7)$ and for $Y_P =0.252$, $N_\nu < 4.4(4.4)$.
(The numbers in parentheses are for the prior $N_\nu \ge 2$.)

\item Using the probability distribution for deuterium
based an extreme model for the chemical evolution of $^3$He,
the 95\% credible region only extends to ${\tilde N}
=3.2$.  For $Y_P=0.242$, $N_\nu<3.4(3.2)$ and for $Y_P=0.252$,
$N_\nu < 3.8(4.0)$.  The use of a probability distribution
for (D/H)$_P$ actually improves the bound slightly,
in spite of the extreme model for chemical evolution assumed.

\item If (D/H)$_P$ is determined to be $(2.5\pm 0.75)\times
10^{-5}$ from high-redshift hydrogen clouds, the 95\% credible limit
region extends to ${\tilde N} = 3.0$.  For $Y_P =
0.242$, $N_\nu < 3.2 (3.0)$ and for $Y_P = 0.252$, $N_\nu < 3.6 (3.6)$.

\item With no information about deuterium,
(D/H)$_P < 1.0$, the likelihood function for ${\tilde N}$ is double-peaked.
The two peaks correspond to two ways of accommodating both
$^4$He and $^7$Li:  low $\tilde N$ and high $\eta$ and high $\tilde N$ and
low $\eta$.  The upper boundary of the 95\% credible region
is ${\tilde N} = 4.3$.  For a central value $Y_P =0.242$,
$N_\nu < 4.3 (4.3)$ and for $Y_P = 0.252$ (four times the
systematic error estimated in Ref.~\cite{olivesteigman}), $N_\nu < 4.9 (4.9)$.

\end{enumerate}

Several points should be noted.  (1)  Because $^7$Li depletion
in pop II halo stars cannot be proved nor disproved at the moment,
our limits are based upon the $^7$Li abundance that gives the
less stringent limit.  Once the issue of depletion is resolved, the limits
may improve by a modest amount.  Our limits are relatively insensitive
to even larger Li depletion.  For example going from a factor of
two depletion to a factor of three depletion
changes the neutrino limit by at most 0.2.
(2)  If the primeval deuterium abundance is
determined to be at the high end of the current range, say (D/H)$_P
= (1.0\pm 0.3) \times 10^{-4}$, the limits are essentially
as in case (1) above.  This is because such a measurement would
pin $\eta$ at very nearly the same value as the D + $^3$He bound.
However, should (D/H)$_P\simeq (2\pm 0.6)\times 10^{-4}$,
the limit worsens slightly (to 4.3 for a central $Y_P =0.242$),
because $\eta$ is pinned to a lower value.
(3) The limits quoted here are not directly comparable to other work.
The most similar approach is that of Fields et al. \cite{recentnulimits};
however, they determine a lower limit to $\eta$ based upon
$^4$He and $^7$Li alone (assuming $N_\nu =3$),
and then use it to set an upper bound to $N_\nu$ (as opposed to
determining a likelihood function for $N_\nu$).

The Table illustrates the robustness of the cosmological
limit to the number of light particle species.  With conservative
assumptions about the astrophysical uncertainties -- the
extreme model for $^3$He chemical evolution and $Y_P = 0.252$ --
less than four massless neutrino species are allowed.  A comparable
limit follows with $Y_P = 0.247\pm 0.003$ and (D/H)$_P < 2\times
10^{-4}$ or no bound to deuterium and $Y_P = 0.242\pm 0.003$.
Even with the most extreme assumptions -- no limit to primeval
deuterium and $Y_P=0.252\pm 0.003$ -- a meaningful limit still
exists, $N_\nu < 4.9$.  Thus, until the current astrophysical
uncertainties are clarified we would argue that $N_\nu <4$ is
a conservative limit, and $N_\nu < 5$ is a very cautious limit.

Astrophysical and cosmological limits will always have
irreducible systematic uncertainty that cannot be quantified
with a standard error.  In the example at hand, the correctness
of the standard model of nucleosynthesis must be assumed -- though
there is no reason to doubt or a compelling alternative --
and abundances in the contemporary Universe must be extrapolated
to their primeval values.  The primary concerns that motivated our
reexamination of the big-bang neutrino limit
are the systematic uncertainties in the primeval D and $^4$He abundances,
with the former dominated at this time by the
chemical evolution of $^3$He.  We have shown that these concerns
do not preclude setting a meaningful and robust limit to $N_\nu$.
With conservative assumptions, consideration of $^7$Li and $^4$He
alone provides the limit, $N_\nu < 4$.  Even with extreme
assumptions, the limit $N_\nu < 5$ follows.  The uncertainty due
to the chemical evolution of $^3$He will vanish when the
primeval abundance of D is determined unambiguously
in high-redshift hydrogen clouds, which, as we have shown, will sharpen
the cosmological limit to the number of light particle species.

\paragraph{Acknowledgments.} 
We acknowledge useful discussions with M. Lemoine, J. Truran, B. Fields
and K. Olive.  This work was supported by the DoE (at Chicago and Fermilab)
and by the NASA (at Fermilab by grant NAG 5-2788 and at Chicago by a GSRP
fellowship).

%%\newpage

%%\section*{Figure Captions}

%%\bigskip
%%\noindent{\bf Figure 1:}  

\begin{figure} \center\leavevmode \epsfxsize=\hsize
\epsfbox{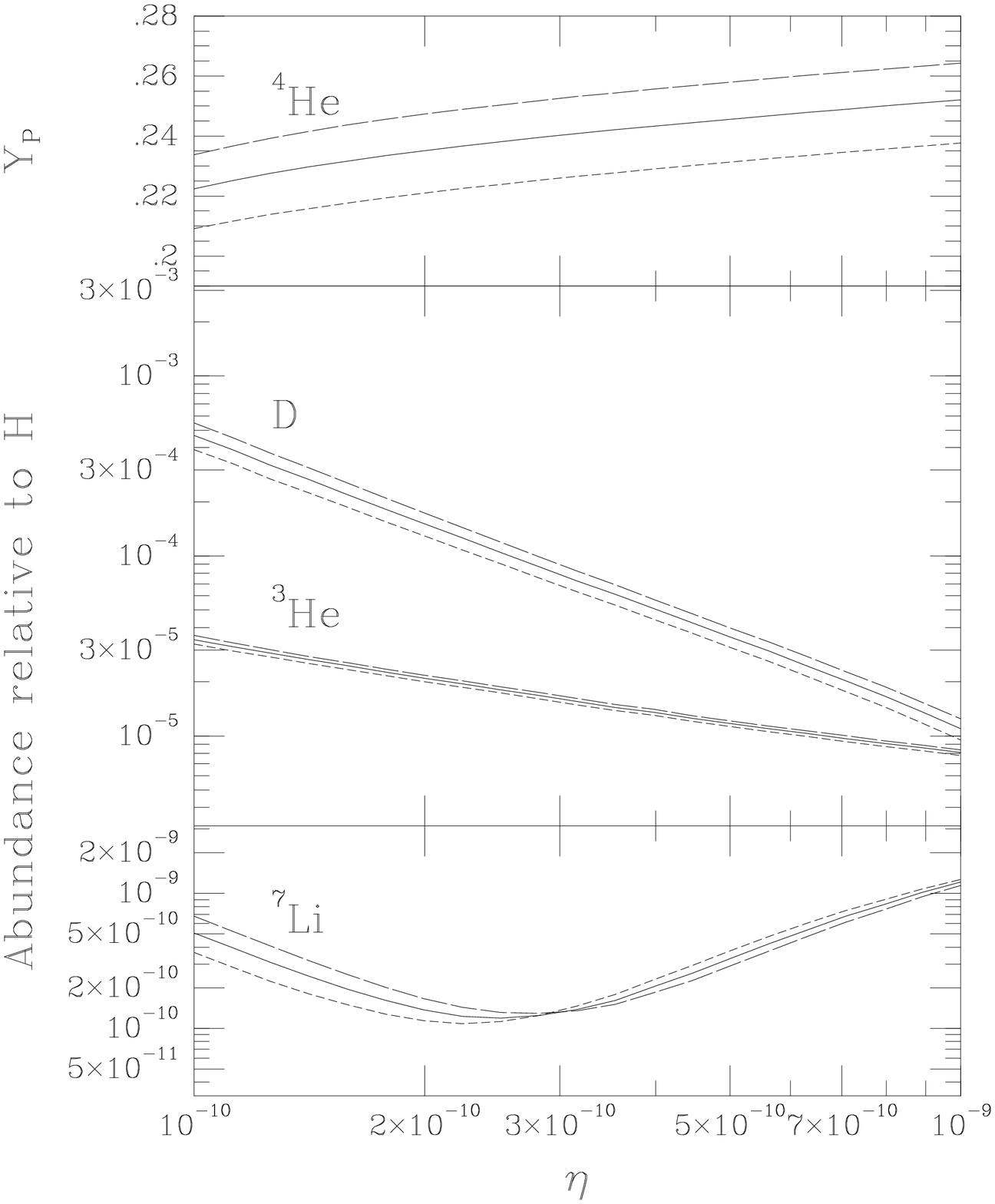}
\caption{Big-bang production of the light
elements for $N_\nu = 2$ (short-dashed line), $N_\nu=3$ (solid line), and
$N_\nu = 4$ (long-dashed line).  Note that for low baryon
density the yields of both $^7$Li and $^4$He increase with $N_\nu$
so that $^7$Li and $^4$He can be used together to
limit $N_\nu$.  (For clarity, the theoretical uncertainty
in the light-element production is not shown; see e.g.,
Ref.~\protect\cite{cst1}.)}
\end{figure}

%%\medskip
%%\noindent{\bf Figure 2:}  

\begin{figure} \center\leavevmode \epsfysize=\hsize
\rotate[r]{\epsfbox{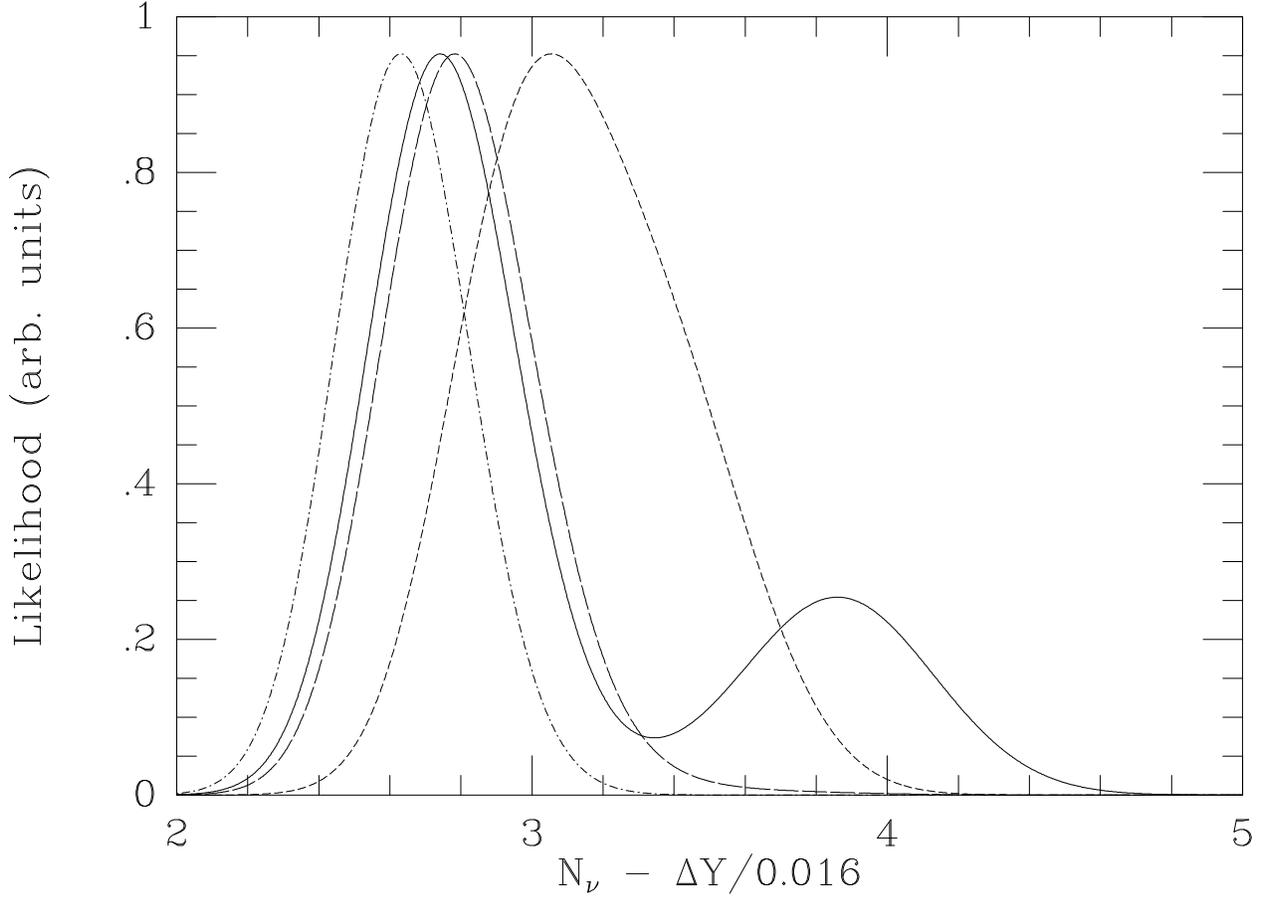}}
\caption{Marginal likelihood for ${\tilde N} \equiv
N_\nu -\Delta Y/0.016$ with different Bayesian priors for the
primeval deuterium abundance:
(D/H)$_P \le 1.0$ (solid line); [(D + $^3$He)/H]$_P \le 2\times 10^{-4}$
(short-dashed line); extreme model of $^3$He chemical evolution
(from Ref.~\protect\cite{ttsc}) (long-dashed line);
(D/H)$_P = (2.5\pm 0.5)\times 10^{-5}$ (dashed-dotted line).
In each case we have assumed the $^7$Li abundance that results
in the least stringent limit to $\tilde N$.  The fact that
$\tilde N = 3$ is well within the 95\% credibility interval
is indicative of the consistency of big-bang nucleosynthesis
with three massless neutrino species.}
\end{figure}

\end{document}